\documentclass[twocolumn,showpacs,preprintnumbers,amsmath,amssymb]{revtex4-1}
\usepackage{graphicx}
\usepackage{color}
\usepackage{braket}
\begin{document}

\title{Ghost Diffraction in Time Domain:\\
Two-Photon Reciprocal Two-Slit Diffraction-Interference}

\author{Yoshiki O-oka}
\affiliation{Graduate School of Arts and Sciences, University of Tokyo, Komaba, Meguro, Tokyo 153-8902, Japan}

\author{Susumu Fukatsu}
\email{To whom all correspondence should be adressed: \\
cfkatz@mail.ecc.u-tokyo.ac.jp}
\affiliation{Graduate School of Arts and Sciences, University of Tokyo, Komaba, Meguro, Tokyo 153-8902, Japan}

\begin{abstract}
\noindent
Temporal ghost diffraction (TGD) is observed by taking two-photon cross-correlation in the configuration of \textit{reciprocal} two-slit diffraction (2SD) where interference fringes develop on the light source side as opposed to the detector side. To this end, a narrow-band chaotic light source and a gated detector are used in the frequency-time domain, which emulate a randomly pointing incoherent light source and a stationary pinhole detector in the momentum-space domain, respectively. Spectral fringes with visibility that closely follows a sinc function of spectral bandwidth are clear evidence that legitimate TGD fringes due to two-photon interference develop even with classical light. 
\end{abstract}


\maketitle
\newpage

Photon correlation has been the target of continued scientific interest for decades from not only the fundamental but application points of view in the context of quantum physics and quantum information processing. Ghost diffraction (GD) is one such unique example that photon correlation is relevant; cross-correlation of photons allows retrieval of diffraction-interference patterns of a remote object without the photon-object interaction.

The GD starts by preparing correlated photon pairs. Then an object (Young's two-slit) is placed in the \textit{test} arm with a stationary detector, from which only the timing of photon arrival is available. The second or \textit{reference} arm has only a space-resolving detector. Although no useful information on the object is available from each arm, cross-correlating sporadic detector readings in both arms can retrieve diffraction-interference fringes of the object \cite{strekalov1995observation}. As such, the GD shares much with ghost imaging (GI) \cite{pittman1995optical, bennink2002two, valencia2005two, ferri2005high, shapiro2008computational, katz2009compressive,ferri2010differential, clemente2010optical, meyers2011turbulence, sun20133d, pelliccia2016experimental, janassek2018ghost}, which allows retrieval of object images using two-photon correlation without directly seeing it. 

Many of the previous GD and GI studies are limited to the space domain\cite{liu2007fourier, hozawa2013single, shi2014polarimetric, jha2015spectral}. Recently, the GI was attempted in the time domain \cite{ryczkowski2016ghost} followed by its differential couterpart\cite{o2017differential}, which reflects the intensive pursuit of time-domain GI analogues in the context of sensing, communication and data processing\cite{shirai2010temporal, chen2013temporal, devaux2016computational, ryczkowski2016ghost, devaux2016temporal, ryczkowski2017magnified, kuusela2017temporal, pan2017optical}. In contrast, viable experimental protocols for implementing the GD in the time domain or temporal GD (TGD) are few\cite{bellini2003nonlocal, viciani2004nonlocal,lajunen2008ghost, shirai2010temporal, setala2010fractional, shirai2017modern}.

The GD is based on \textit{pinhole} detection while the GI is based on \textit{bucket} detection. Such a discrepancy was noticed from the beginning\cite{strekalov1995observation}, but its physical significance has obscured over time. This is because the focus was more on the relevance of nonclassicality of the input photons and two-photon interference (TPI) although subsequent GD attempts using classical light dismissed the nonclassicality issue\cite{ferri2005high, zhang2007lensless, yu2016fourier}. Importantly, it was later pointed out that the GD is fully accounted for by the theory of propagating light waves using transfer functions \cite{gatti2003entangled, gatti2004ghost}. Indeed it explains why ghost diffraction-interference fringes develop in the reference arm using the $f-f$ lens optics \textit{without} addressing the specific role of pinhole detector, which has even discouraged discussion on the relevance of the TPI in the GD\cite{strekalov1995observation}.  Thus the significance of pinhole detection and TPI remains barely understood. On the other hand, such a notion has lead to the difficulty achieving TGD due mainly to the lack of temporal lens optics\cite{shirai2010temporal}, which is further complicated by the fact that a 1-D time-domain analog of GI-GD transformation in the 3-D space-domain is not trivial.

In this Letter, we report the first TGD by implementing \textit{reciprocal} two-slit diffraction (2SD) in the frequency-time domain. The reciprocal 2SD not only allows transformation of the GD geometry in the space domain into one in the time domain, but also clarify the significance of pinhole detection and the relevance of the TPI for more generic cases using classical light.  

\begin{figure}[!t]
\begin{center}
\includegraphics[width=69mm, bb=0 0 260 330]{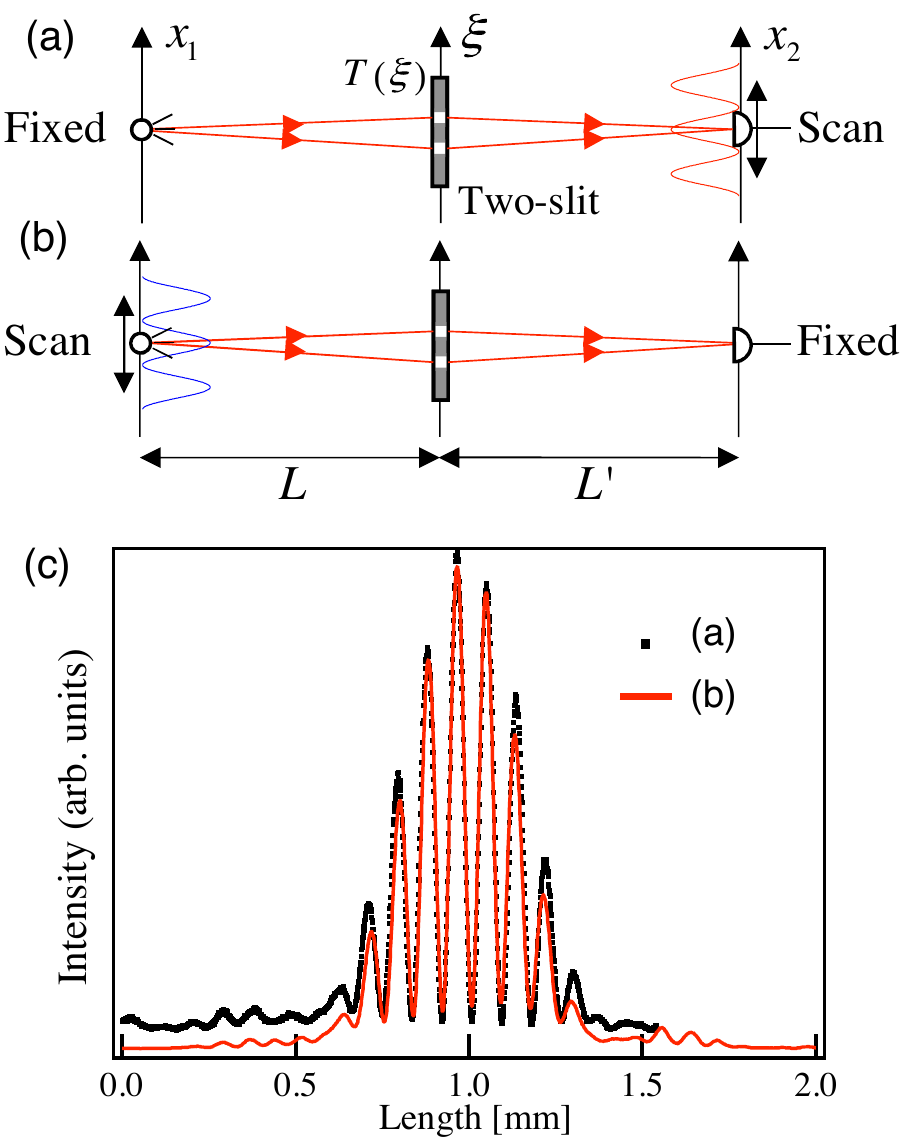}
\end{center}
\caption{(a) \textit{One-photon} two-slit diffraction (2SD) and (b) one-photon reciprocal 2SD. Interference fringes develop in 2SD as the detector scans over the screen with the point source fixed. Reciprocal 2SD allows interference fringes to appear on the light source side as the point source moves while the detector is fixed. (c) Measured diffraction-interference fringes of 2SD (dotted) and reciprocal 2SD (solid) for $L=L^{\prime}$.}
\label{Fig.1}
\end{figure}

Figure \ref{Fig.1}(a) shows the schematic of \textit{one-photon} 2SD. The electric field $E$ at the position $x_{2}$ on the screen parallel to the slit is given by the Fourier transform, $\mathcal{F}$, 
\begin{equation}
\small{
E\left(x_2, 0, L^{\prime}\right)
=E_0\Theta \mathcal{F_{\xi}}[T(\xi, 0, 0)]\left[k\left(\frac{x_1}{L}+\frac{x_2}{L'}\right)\right] }
\label{eq1}
\end{equation}
where $E_0$ is the amplitude, $x_{1}$ is the $x$-component of the coordinate $(x_1, y_1,  -L)$ in the light-source plane, $(\xi, \eta, 0)$ is the slit-plane coordinate referred to the center of the slits, $L$($L^{\prime}$) is the source-slit(slit-detector) distance, $k$ is the free-space wavenumber of light, and $\Theta$ is the phase factor that depends on $k$, $L$, $L^{\prime}$, $x_1$ and $x_2$.

The symmetry of Eq.\ (\ref{eq1}) with respect to $x_{2}/L^{\prime}$ and $x_{1}/L$ reveals \textit{reciprocity} in the 2SD geometry: if electric fields from a hypothetical point source at the detector position, $x_{2}$, passed through the two-slit, diffraction-interference patterns would develop \textit{on the light source side}. This is confirmable by performing the \textit{reciprocal} 2SD experiment sketched in Fig.\ \ref{Fig.1}(b). To do so, the detector position $x_{2}$ is fixed while the light source is scanned to see if the interference fringes $I(x_1)$ develop such that

\begin{equation}
\small{
I(x_1)\equiv I(x_1, 0, -L)=2I_0\left[1+\cos \left(k\frac{x_1}{L}d+k\frac{x_2}{L'}d \right)\right]}
\label{eq2}
\end{equation}
where $d$ is the slit separation and $I_0$ is a positive constant. The solid line in Fig.\ \ref{Fig.1} (c) shows the result of \textit{reciprocal} 2SD with $L=L^{\prime}$. The incoherent light source is a HeNe ion laser with a 50-$\mu$m circular aperture producing diffracted randomly-pointing photons. The detector is a photon-counting device with $25\times25 \mu$m$^2$ aperture. Apparently, a close match is found between the solid line (red) and the interference fringes due to normal 2SD shown by the dotted line (black).

\begin{figure}[!t]
\begin{center}
\includegraphics[width=80mm, bb=0 0 400 430]{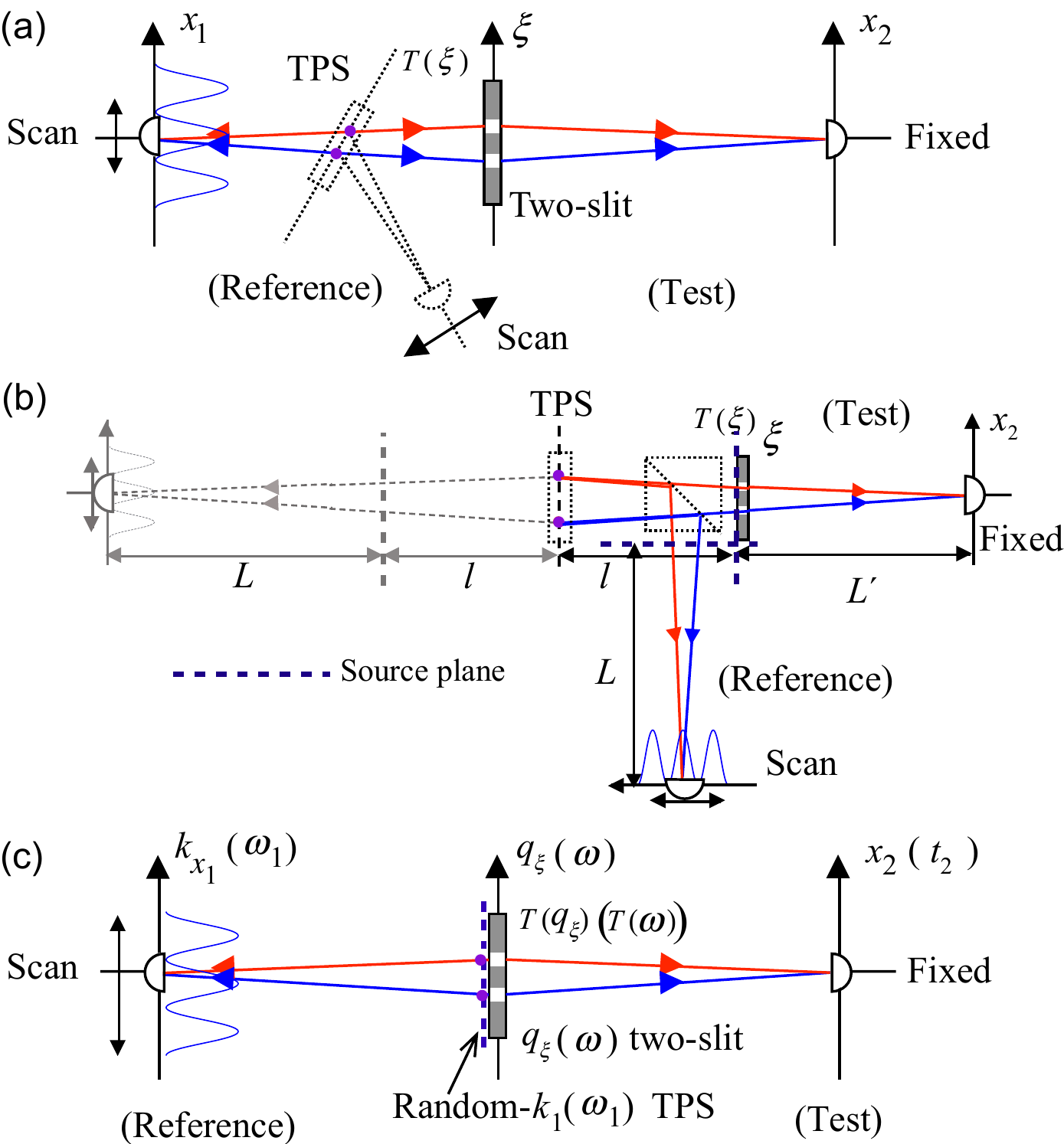}
\end{center}
 \caption{(a) \textit{Two-photon} reciprocal two-slit diffraction (2SD) using two detectors and an incoherent two-photon source (TPS) placed in the light paths (dotted lines). Folding the left part with respect to TPS reproduces the GD geometry. (b) A variant of (a) using a beam splitter. Source planes are indicated by thick broken lines. Reference arm can be unfolded (gray lines on the left of TPS) to reproduce a configuration equivalent to (a). (c) Schematic GD diagram using conjugate variables in the space (time) domain based on two-photon reciprocal 2SD. See text for details. Note that two-photon paths are indicated by red and blue lines with arrows. }
\label{Fig.2}
\end{figure}

This has the following implications. First of all, the pinhole detector is now given the role of the hypothetical light source, which explains why it must be \textit{point-like}.  Second, which-path information must be erased on \textit{both sides} of the slit for interferences to occur. Last but significant, device functions do not count, so "source" and "detector" are \textit{interchangeable}. Thus one can design a \textit{two-photon} reciprocal 2SD setup with two detectors. In this case, however, an incoherent light source emitting photons on either side must be placed somewhere in the light paths in Fig.\ \ref{Fig.2}(a) (dotted rectangle), which emulates the configuration of the first GD attempt \cite{strekalov1995observation}. If the left arm with a second detector (reference arm) is folded back with respect to the light source onto the side of the arm with the original detector and the two-slit (test arm) (dotted line in Fig.\ \ref{Fig.2}(a)), the generic TPI geometry is obtained such that the photon stream from an incoherent light source is split along the test and reference arms.

Here we consider a more general case using classical light and a beam splitter (BS) as opposed to Ref.\ 1\ (Fig.\ \ref{Fig.2}(b)). As such, the TPI due to the \textit{bunched} two-photon state $\ket{2}\!\ket{0}+\ket{0}\!\ket{2}$ (indicated by the blue and red lines) rather than the \textit{separable} one $\ket{1}\!\ket{1}$ is relevant, where the kets refer to the distinguishable or separated emitter positions\cite{strekalov1995observation}. We have that

\begin{eqnarray}
\small{\langle I_{\mathrm t}(x_2)I_{\mathrm r}(x_1)\rangle=2I_0^2\left[1+\cos k\left(\frac{x_1}{L}+\frac{x_2}{L^{\prime}}\right)d\right]}
\label{eq3}
\end{eqnarray}
where $I_0$ is a constant, $L$ is the distance between the BS and the detector in the reference arm, and $L^{\prime}$ is the distance between the slit and the screen. Equation (\ref{eq3}) provides the intensity or second-order fringes expected for GD as a function of $x_1$ when $x_2$ is fixed. On the other hand, the first-order fringes vanish in the test arm,
\begin{equation}
\small{\langle I_{\rm t}(x_2)\rangle_{x_1}=2I_0\left[1+\langle\cos k\left(\frac{x_1}{L}+\frac{x_2}{L^{\prime}}\right)d\rangle_{x_1}\right]=2I_0}.
\label{eq4}
\end{equation}
This reminds us of what happens to an entangled two-photon state where only joint detection allows interferences to occur while it ends up mixed otherwise. With all the features in the first GD experiment reproduced\cite{strekalov1995observation}, the previous GD attempts seem to be consistently explained in terms of reciprocal 2SD where the TPI is relevant. 

To apply such a reciprocal 2SD protocol to the time domain, we use conjugate variables in the Fourier space  (Figure \ref{Fig.2}(c)). In normal 2SD, Eq.\ (\ref{eq1}) is written as 
\begin{equation}
E(x_1, 0, -L)=E_0\Theta^{\prime} 
\int d\xi E(\xi, 0, 0)T(\xi)e^{-ik\frac{x_1}{L}\xi}.
\label{eq5}
\end{equation}
where $\Theta^{\prime}$ is the phase factor. Note here that $k\frac{x_1}{L}$ is the in-plane wavevector along the screen since we find that

\begin{equation}
k\frac{x_1}{L}\approx k \times \frac{x_1}{\sqrt{x_1^2+L^2}} \approx k \times \frac{x_1-\xi}{\sqrt{(x_1-\xi)^2+L^2}}  \equiv k_{x_1}. 
\label{eq6}
\end{equation}
Seen from the detector at $x_2$, the integration with respect to $\xi$ (Eq.\ (\ref{eq5})) is convertible to the one with respect to the wavevector $q_{\xi}$ lying in the slit plane for $|\xi|< |x_2| \ll L^{\prime}$, 
\begin{equation}
q_{\xi} \equiv  k_{\xi}(x_2,\xi)-k_{\xi}(x_2,0) \approx k \times \frac{\xi}{\sqrt{x_2^2+{L^{\prime}}^2}}\approx k\frac{\xi}{L^{\prime}}.
\label{eq7}
\end{equation}
Substituting $q_{\xi}=\pm \frac{1}{2}k\frac{d}{L^{\prime}}$ into Eq.\ (\ref{eq5}) yields 

\begin{equation}
E(x_1, 0, -L)=E_0\Theta  \left\{2\cos\left(k_{x_1}\frac{d}{2}+\frac{kd}{2L^{\prime}}x_2\right)\right\}.
\label{eq8}
\end{equation}

\begin{figure}[!t]
\begin{center}
\includegraphics[width=75mm, bb=0 0 500 360]{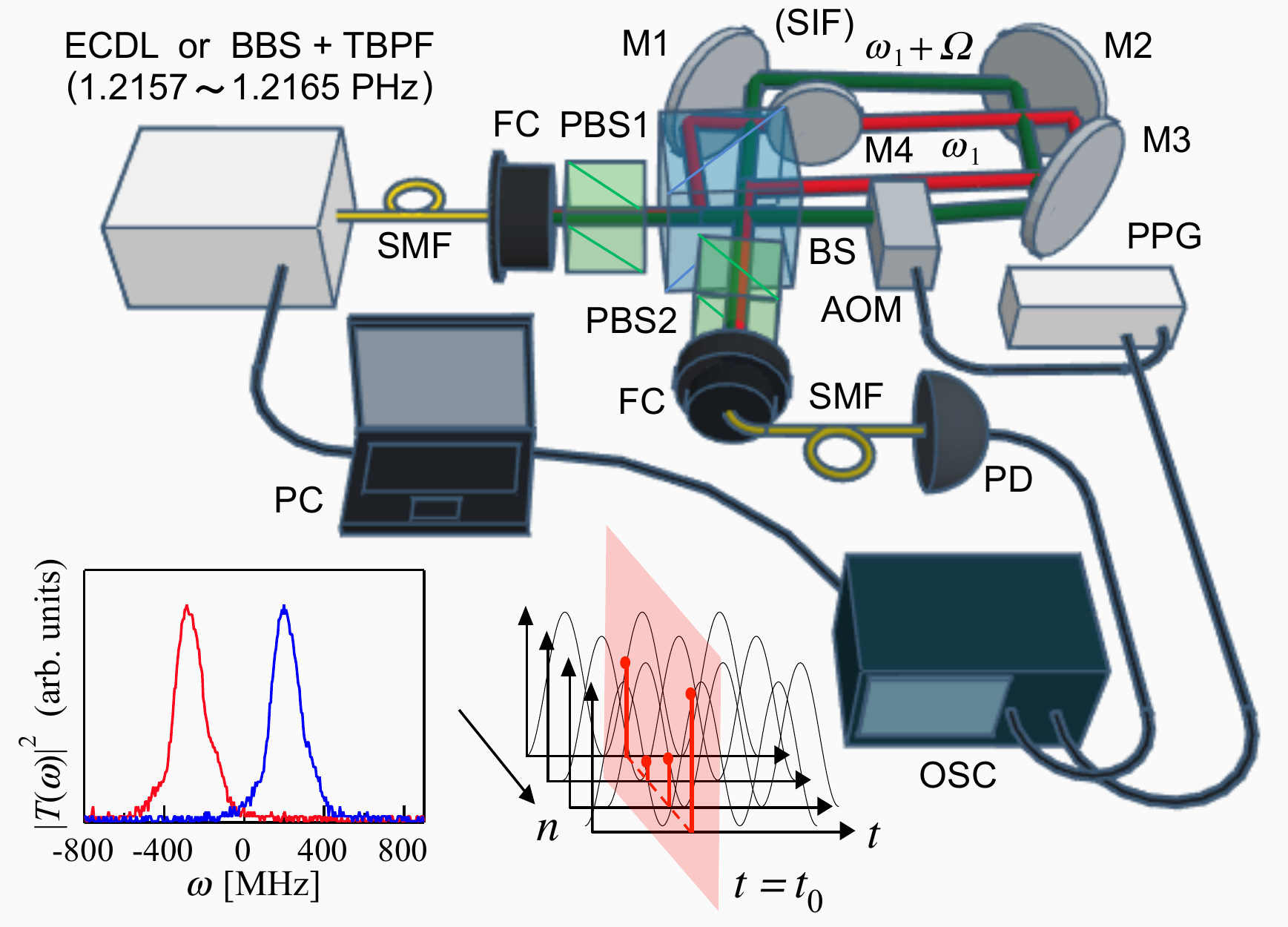}
\end{center}
\caption{Schematic of computational temporal ghost diffraction (TGD). ECDL, external cavity diode laser; BBS, broadband source; TBPF, tunable bandpass filter; SIF, Sagnac interferometer; BS, non-polarizing beam splitter; PBS, polarization beam splitter; AOM, acousto-optic modulator; PD, photodiode; SMF, single-mode fiber; PPG, pulse pattern generator; OSC, oscilloscope. Left inset: spectra of the fundamental $\omega$ and its side band, $\omega+\Omega$. Right inset: TGD interference fringes are evidenced by the spectral profile in the cross-section (shaded) of heterodyne beats at local time $t_0$.  }
\label{Fig.3}
\end{figure}

\begin{figure}[!t]
\begin{center}
\includegraphics[width=70mm, bb=0 0 550 600]{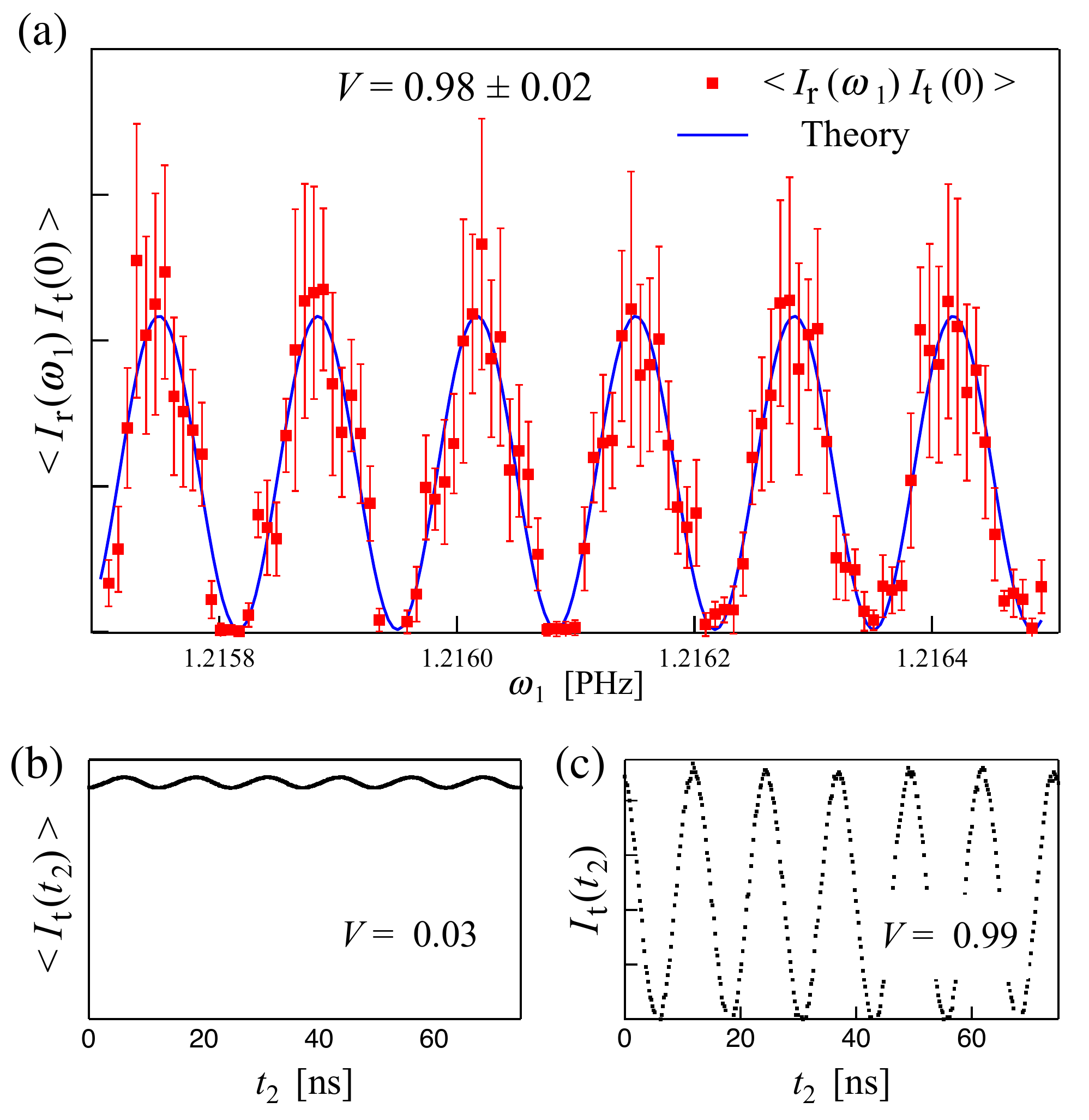}
\end{center}
\caption{(a) Intensity cross-correlation, $\langle I_{\mathrm r} \left(\omega_{1} \right) I_{\mathrm t}\left(0\right) \rangle$, versus angular frequency, $\omega_{1}$. Spectral fringes are seen with visibility $V$=0.98$\pm$0.02. (b) First-order interference fringes in the test arm. The remaining oscillations result from averaging over a finite $\omega_{1}$-interval of (a). (c) Heterodyne beat for a fixed $\omega_{1}$.}
\label{Fig.4}
\end{figure}

In essence, the two-photon reciprocal 2SD emulates the first-order diffraction-interference of light using the TPI geometry (Fig.\ \ref{Fig.2}(a)). So it turns out from Eq.\ (\ref{eq8}) that the GD is the process of \textit{post-selecting} $k_{x_1}$ through two-photon cross-correlation $\langle I_{\mathrm t}(x_2) I_{\mathrm r}(k_{x_1})\rangle$ between the fixed detector ($x_2$) in the test arm and that in the reference arm as $k_{x_1}$ varies with the scanning detector. In practice, one chooses the detector position in the test arm somewhere in interference fringes (albeit invisible before post-selection), which \textit{presets} a phase to be post-selected. Then $k_{x_1}$ is varied in the reference arm, and only those $k_{x_1}$'s which are \textit{coherent} or in phase with the preset phase are kept. Now we design a TGD experiment with reference to Fig.\ \ref{Fig.2}(c). In the frequency domain, $k_{x_1}$ reads $\omega_1$ and the phase is preset by selecting the detector position, $t_2=t_0$, somewhere on heterodyne beats, viz. temporal diffraction-interference fringes. As a frequency analog of two-slit, we prepare light fields oscillating at $(\omega_1+\Omega/2)\pm\Omega/2$ with a frequency shifter which up-converts the angular frequency of the incoming light fields by $\Omega$. As $\omega_1$ is randomly scanned, only data coherent with the preset phase are post-selected, which eventually yields spectral fringes, $\langle I_{\mathrm t}(t_0) I_{\mathrm r}(\omega_1)\rangle$, as a signature of the TGD.

Figure \ref{Fig.3} shows the schematic TGD setup. The light source is a 7-dBm external-cavity diode laser with bandwidth $<$ 1 MHz. The reference arm is taken over by a set of randomized frequency data, $I_{\mathrm r}(\omega_1)$, in the range 1.2157-1.2165 PHz, issued from a laptop, which makes our experiment essentially \textit{computational} TGD. In the test arm, a non-polarizing beam splitter at the input of the Sagnac interferometer (SFI) splits the incoming light into two propagation modes, which are eventually merged and coupled into a single-mode fiber leading to a 5-GHz detector. An acousto-optic modulator (AOM) driven at $\Omega=$ 502 MHz by a pulse pattern generator (PPG) is placed in one of the light paths. The fundamental ($\omega_1$) and its sideband ($\omega_1+\Omega$) spectra are visible in the left inset of Fig.\ \ref{Fig.3}. Heterodyne beats are captured on an oscilloscope triggered by the PPG. The data acquired in the test arm at \textit{local} time $t_0$ in the capture frame of heterodyne beats, $I_{\mathrm t}(t_2=t_0)$, are cross-correlated with $I_{\mathrm r}(\omega_1)$, which are averaged over an ensemble of $10^3$ such that

\begin{equation}
\small
\langle I_{\mathrm r} \left(\omega_{1} \right) I_{\mathrm t}\left(t_0\right) \rangle=2I_0^2\left[ 1+{\rm cos}\left( \omega_{1}\frac{\mathit{\Delta} r}{c} + \Omega t_0 +\varphi \right)\right] 
\label{eq9}
\end{equation}
where $\Delta r$ is the delay in the interferometer and $c$ is the speed of light. $\varphi$ is the extra phase dependent on the modulation frequency, $\Omega$, in such a way that $\varphi =\Omega/c r_{\mathrm b}$ where $r_{\mathrm b}$ is the distance between the SFI exit and AOM, and $\Delta r=r_{1}-r_{\mathrm a}-r_{\mathrm b}$ where $r_{1}$ is the arm length of the SFI while $r_{\mathrm a}$ is the distance between the SFI entrance and AOM. A separate experiment using monochromatic light at $\omega_1=$1.2161 PHz has shown that $\Delta r$= 14 mm.  

\begin{figure}[!t]
\begin{center}
\includegraphics[width=70mm, bb=0 0 420 709]{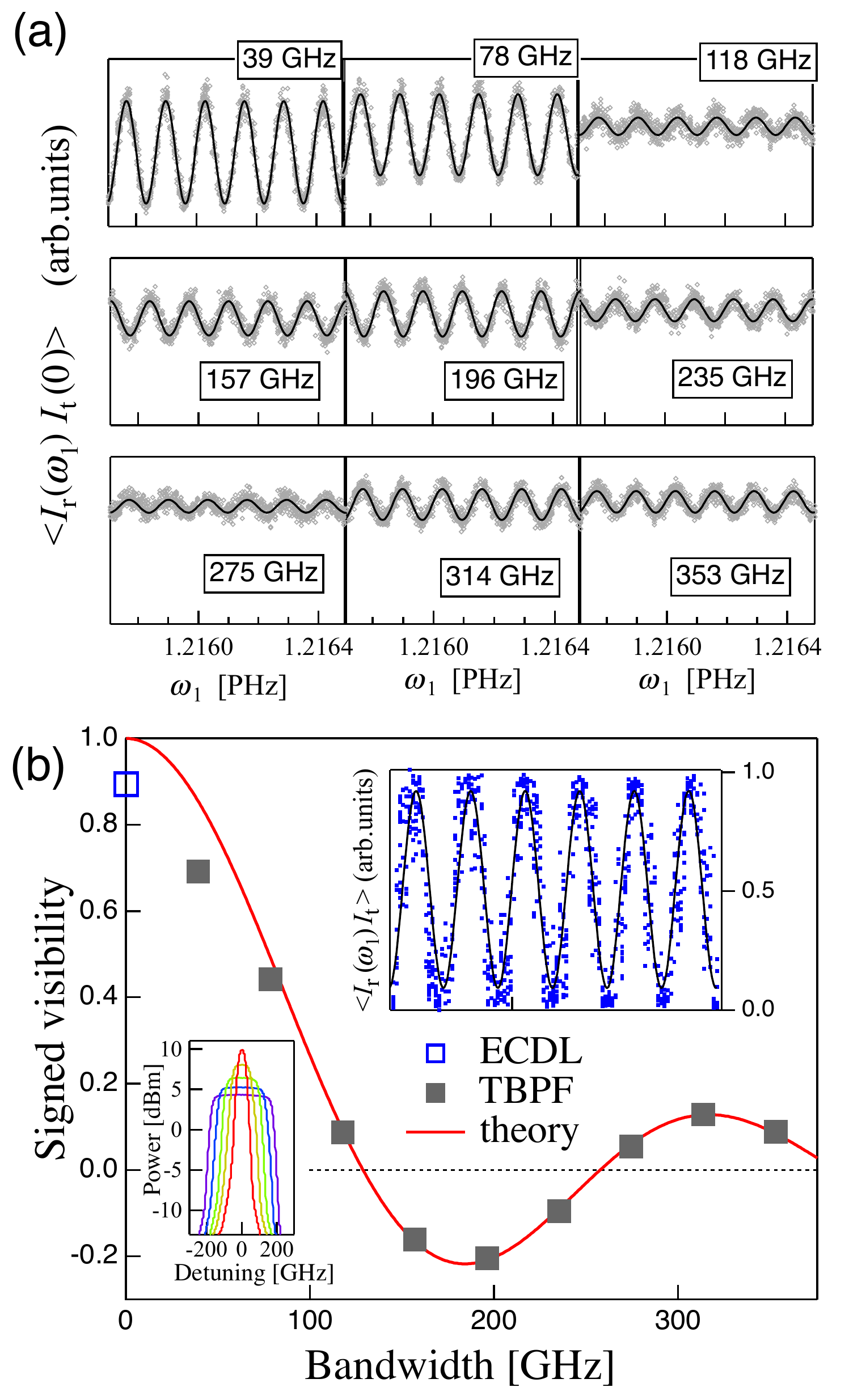}
\end{center}
\caption{(a) $\langle I_{\mathrm r} \left(\omega_{1} \right) I_{\mathrm t}\left(0\right)  \rangle$ as a function of spectral bandwidth, $\Delta\omega_{1}$. Solid lines are theoretical fits. Note a $\pi$-phase shift between the rows. (b) Signed visibility ($V$) plotted against $\Delta\omega_{1} $. Solid line is the sinc function of Eq. (\ref{eq10}).  Upper inset figure shows spectral fringes obtained by using the external-cavity diode laser. Lower inset shows the typical filtered spectra.}
\label{Fig.5}
\end{figure}

Figure \ref{Fig.4}(a) shows the measured intensity cross-correlation $\langle I_{\mathrm r} \left(\omega_{1} \right) I_{\mathrm t}\left(t_2=0\right)\rangle$, i.e., spectral fringes. The solid line is the theoretical fit according to Eq.\ (\ref{eq9}), which yields a high visibility $V= 0.98\pm 0.02$. This is compelling evidence that classical light allows the TGD by means of two-photon reciprocal 2SD. Shown in Fig. \ref{Fig.4}(b) for comparative purposes is the one-photon correlation $\langle I_{\mathrm t}(t_2)\rangle$ with only minor fringes with $V\approx0.03$. Figure  \ref{Fig.4}(c) shows a typical trace of heterodyne beats with $V\approx 0.99$, from which spectral fringes in Fig. \ref{Fig.4}(a) are to be retrieved.

Finally, the spectral bandwidth ($\Delta\omega_1$) (or "slit width") dependence is studied as a stringent check on the legitimacy of the TGD. The light source is a tunable bandpass filter  (TBPF) with 10$^{3}$-dB/nm roll-off (Santec OTF-980) inserted between two 15-dBm erbium-doped fiber amplifiers. The one before TBPF is the broadband source utilizing amplified spontaneous emission (ASE). Note that $\omega_1=$1.2161 PHz and $\Delta\omega_1$ = 39-353 GHz. The panels in Fig. \ref{Fig.5}(a) show the cross-correlation $\langle I_{\mathrm r} \left( \omega_1\right) I_{\mathrm t}\left(0\right) \rangle$ as a function of $\Delta\omega_1$  that must follow, if the TGD is relevant,
\begin{eqnarray}
\lefteqn {\langle I_{\mathrm r} \left(\omega_1 \right) I_{\mathrm t}\left(t_2\right) \rangle_{\Delta\omega_1}=} \nonumber\\
&&2I_0^2\left[ 1+{\rm sinc}\left(\frac{\mathit{\Delta} r}{2c} \Delta\omega_1\right){\rm cos}\left(\frac{\mathit{\Delta} r}{c}\omega_1+\Omega t_2 +\varphi \right)\right].
\label{eq10}
\end{eqnarray}
We define the \textit{signed} visibility, $\tilde V={\rm sinc}\left(\frac{\mathit{\Delta} r}{2c} \Delta\omega_1\right)$, which has the magnitude $V=\left|{\rm sinc}\left(\frac{\mathit{\Delta} r}{2c}\Delta\omega_1\right)\right|$ and the phase, 0 or $\pi$. In fact, phase reversals between the rows are visible. Fitting using Eq. (\ref{eq10}) with $\Delta r$=14 mm yields the traces shown by the solid lines. The $\tilde V$ values are plotted in Fig. \ref{Fig.5}(b) versus $\Delta\omega_1$. The solid line is the sinc part of Eq. (\ref{eq10}). Clearly, the experimental data match the theory for $\Delta\omega_1 \ge$ 118 GHz. A slight departure for  $\Delta\omega_1 \le$ 78 GHz indicates the relevance of the background ASE giving a positive offset to Eq. (\ref{eq10}) and hence a lower $V$ values. 

Although our TGD builds on \textit{heterodyning}, it is tempting to put $\Omega=0$ so that $\langle I_{\rm r}\left( \omega_1 \right) I_{\rm t} \rangle$ reduces to \textit{homodyning} where the path length difference $\Delta r$ is concerned. This allows for a different class of time-domain GD using a \textit{temporal} two-slit. In Refs.\ 27 and 28, spectral fringes observed by implementing a Michelson interferometer in the test arm are claimed to be a manifestation of the TGD using such a two-slit. However, they used a \textit{bucket} detector instead of a pinhole detector, which makes us believe that the observed spectral fringes are \textit{ghost images of the spectra} rather than ghost diffraction-interference fringes. Most importantly, there is a profound difference in the intensity correlation; Briefly, theirs should be written as $1+\cos\omega_{1} t_2$, so $\omega_{1}$ and $t_2$ are inseparable unlike ours, $1+\cos(\alpha\omega_{1}+\beta t_2)$, where $\alpha$ and $\beta$ are constants.  

In summary, temporal ghost diffraction (TGD) was demonstrated by computationally taking time-gated cross-correlation of frequency-randomized photons in a \textit{reciprocal} two-slit diffraction configuration using classical light. Spectral fringes with bandwidth-dependent signed visibility were clearly observed in the two-photon interferometer geometry, which is evidence for the TGD. 
	
The authors acknowledge the technical support from Y. Yasutake. This work was in part supported by JSPS KAKENHI 16K13714 and 17H02773.

\bibliography{GI_reference.bib}

\end{document}